\documentclass[12pt,preprint]{aastex}

\slugcomment{18/11/2002, in HKU}

\shorttitle{An accretion model for the growth of the central black
hole....} \shortauthors{Lu et al.}

\begin{document}


\title{An accretion model for the growth of the central black hole associated with ionization instability in quasars}
\author{Y. Lu \altaffilmark{1}}
\affil{National Astronomical Observatories, Chinese Academy of
Sciences, Beijing, China, 100012 } \email{ly@bao.ac.cn}

\author{K.S.Cheng}
\affil{Department of Physics, The University of Hong Kong, Hong
Kong, China}

\and

\author{S.N.Zhang\altaffilmark{2}}
\affil{Physics Department and Center for Astrophysics, Tsinghua
University, Beijing, 100084, China }

\altaffiltext{1}{Visiting scholar, Department of Physics, The
University of Hong Kong, Hong Kong, China} \altaffiltext{2}{Also
Physics Department, University of Alabama in Huntsville,
Huntsville, AL 35899, USA}


\begin{abstract}
 A possible accretion
model associated with the ionization instability of quasar disks
is proposed to address the growth of the central black hole
harbored in the host galaxy. The evolution of quasars in cosmic
time is assumed to change from a highly active state to a
quiescent state triggered by the S-shaped ionization instability
of the quasar accretion disk. For a given external mass transfer
rate $\dot{M}_{ext}$ supplied by the quasar host galaxy,
ionization instability can modify accretion rate in the disk and
separates the accretion flows of the disk into three different
phases, like a S-shape. We suggest that the bright quasars
observed today are those quasars with disks in the upper branch of
S-shaped instability, and the faint or 'dormant' quasars are
simply the system in the lower branch. The middle branch is the
transition state which is unstable. We assume the quasar disk
evolves according to the advection-dominated inflow-outflow
solutions (ADIOS) configuration in the stable lower branch of
S-shaped instability, and Eddington accretion rate is used to
constrain the accretion rate in each phase. The mass ratio between
black hole and its host galactic bulge is a nature consequence of
ADIOS. Our model also demonstrates that a seed BH ($\sim 2\times
10^6M_\odot$) similar to those found in spiral galaxies today is
needed to produce a BH with a final mass ($\sim 2\times
10^8M_\odot$).

\end{abstract}

\section{Introduction}
Nuclear feeding is most important in the evolution of quasars and
quasar host galaxies. Since quasars were first discovered, it has
been suggested to be powered by the accretion of gas on to
super-massive black holes (SMBHs) at the center of galaxies
\citep{Lyn69}. In the framework of the hierarchical
dark-matter(DM) cosmology \citep{Hae98}, the formation and
evolution of galaxies and their active nuclei are intimately
related \citep{San96, Boy98, Fab99, Gra01, Hae00}. Consequently,
the growth and the fuelling of SMBH, and the construction of the
galactic bulge, are now seen as being potentially related
processes \citep{Sil98, Fab99, Gra01}. Sub-millimeter photometry
of eight X-ray-absorbed active galactic nuclei (AGNs)\citep{Pag01}
reveals that active nuclei and the galaxy evolve together, with
the black hole (BH) accreting matter and the galaxy creating
stars. On the other hand, recent high-resolution observations of
galactic centers have revealed that the estimated mass of a
central "mass dark object" (MDO), which is the nomenclature for
SMBH candidate, correlates with the mass of a galactic bulge. The
BH - bulge mass relation with $M_{bh}\sim 0.001-0.006M_{bulge}$
\citep{Kor95, Mag98, Mar99, Geb00, Fer00} for normal galaxies was
shown to extend to nearby quasar host galaxies as well
\citep{Lao98, Nel00, Fer01}. Based on the remarkable dependence of
black hole mass on spheroid velocity dispersion \citep{Fer00,
Mer01, Geb00, geb00}, several theoretical models for BH growth
have been considered to explain the BH-to-bulge correlations,
e.g., a hydrodynamic model, including wind regulation
\citep{Sil98}, an inside-out accretion model \citep{Ada01} and a
self-interacting dark matter model \citep{Ost00} or star formation
regulated model \citep{Bur01}. As an alternative model for BH
growth, \cite{Ume01} suggested a radiation-hydrodynamical model
incorporating the physics of angular momentum transfer. The growth
history of the super-massive black holes (BHs) is, however, rather
uncertain. Two scenarios are involved in modeling these BHs: One
is that SMBHs can grow out of low-mass "seed" BHs through
accretion \citep{Ree84}, and another is that SMBHs grow by merging
\citep{Bar00}.

It is widely believed that SMBHs built up during the quasar epoch,
around $z\sim 2-3$, where $z$ is the red-shift of quasars. However
the observation of optically bright quasars shows that if the
quasars were continuously radiating, they would grow too large and
run out of fuel \citep{Hae98, Ric98}. On the other hand, if the
quasar disk were unstable, then quasars would be intermittent, and
mostly non-radiating, with the corresponding long periods of
quiescence resolving the old fuelling problem in quasars
\citep{Shi78}. This implies that the growth of BHs through
accreting gas only happens intermittently \citep{Sma92}.

The S-shaped ionization instability first found to be responsible
for the outbursts of cataclysmic variables
(CVs)\citep{Mey84,Sma84}, also operates in binary systems
containing accreting neutron stars or BHs. The time evolution of
the limited cyclic behavior is caused by the thermal and viscous
instability of the disk  in the region of partial ionization of
the hydrogen and leads to transient behavior of many sources on a
time-scale of years \citep{Kin95,Can93, Ham98}. It is suggested
that the similar instability may occur in the quasar disk
\citep{Lin86, Can92,Min90,Cla89,Sie96,Hat01}, the expected
time-scale being of thousands to millions of years. Such
instability may modify the accretion rate in the central part of
the flow and thus be responsible for a temporary change of the
object status from highly active to quiescent on similar
time-scales. \cite{Bur98} have summarized a decade of work on
quasar disk instabilities and conclude that all observed active
galactic nuclei (AGN) are in the outburst phase. For a central
black hole with mass of $10^8M_\odot$, the recurrence time scale
of the S-shaped instability is about $10^6{\rm yr}$, which the
quasar disk spends about $10\%$ of the time in outburst phase and
$90\%$ in the quiescent phase \citep{Sie99}. It is thus suggested
that the bright quasars we observe today are in a highly active
state and the faint or "dormant" systems are simply those in
quiescent states.

Motivated by the above investigations, we here develop an
alternative accretion model associated with the S-shaped
instability of the quasar disks for the growth of BHs harbored in
quasar host galaxies. The mass relation between BH and the
galactic bulge in which the quasar lies is also discussed.
Contrary to the former BH growth model through accretion
\citep{Ree84}, we address the growth of the BHs using the
ionization instability of the quasar disk. Eddington accretion
rate is used to limit our calculations. The behavior of the disk
is successively "drought", "inundation" and "deluge" \citep{Bla02}
in the lower, middle and upper branches of the quasar disk,
respectively. This paper is organized as follows: in section 2, we
discuss the accretion features of the quasar disk around a BH in
different phases, and study the growth of the BH with the
evolution of the quasar disk; the BH - bulge mass relation is
discussed in section 3; finally, we give our discussion and
conclusion in section 4.

\section{The growth of BHs associated with the S-shaped ionization instability of quasars}
We first assume that there is a seed BH in the center of a quasar
host galaxy with a mass much less than its present day mass, for
example, a seed HB could be formed from an immediate coalescence
of the small-mass BH contained in the merging galaxies
\citep{Beg80, Cat99} or an early formed massive BH with
$10^6M_\odot$ \citep{Ume93}. A massive BH with $\sim 10^5M_\odot$
can also be formed by the collapse of a rotating supper-massive
star \citep{Bau99,Sai02, Car84, Nak01}. The seed BH will grow as
long as there is a gas supply for the inner disk. This gas
reservoir is replenished by viscous inflow from the outer disk
into the inner region of the disk and augmented by the increase in
the inner radius due to the growth in the BH mass. A sufficient
cold gas with transferring rate $\dot{M}_{ext}$ could be supplied
from the quasar host galaxy, for instance, mass loss from stars,
stellar disruption and collisions etc., if the seed BH is embedded
in densely populated stellar systems \citep{Tsu77}; and from the
progenitor galaxies if the seed BH formed in major mergers
\citep{Kau00}. According to the model (i) of \cite{Cat99} that
$\dot{M}_{ext}$ is proportional to the mass that has formed stars
in the host galaxy. We take $\dot{M}_{ext}$ as a model parameter
in this paper. If there is lack of any limiting factor
\citep{Bur01}, all the gas supply will eventually be accreted and
fall into the BH. Within the framework of our model, Eddington
limit \citep{Sma92} is used for the growth of BH, and the
ionization instability of the quasar disk provides the
self-regulation that limits the growth of the central BH in three
different phases.

\subsection{The S-shaped ionization instability of the quasar disk}

The most realistic scenario for an accretion disk around a BH is
time dependent. The stationary local disk solutions form a
characteristic S-shape on the $\dot{M}-\Sigma$ (accretion rate vs.
surface density) plane. The accretion disk is thermally and
viscously unstable when $\frac{\partial\dot{M}}{\partial
\Sigma}<0$ \citep{Mey84}. This is the case for accretion rates in
the range $(\dot{M}_1, \dot{M}_2)$, where the S-shaped has a
negative slope ({ Fig.1}). A characteristic S-shaped instability
consists of three branches (lower, middle, and upper branches)
creating three possible accretion phases of the quasar disk. In
the case of Keplerian motion the surface density evolution may be
written \citep{Pri81, Sma84} as
\begin{eqnarray}
\frac{\partial\Sigma}{\partial
t}&=&\frac{3}{r}\frac{\partial}{\partial
r}\left[r^{1/2}\frac{\partial}{\partial r}(\nu\Sigma
r^{/2})\right]+\frac{1}{2\pi r}\frac{\partial\dot{M}}{\partial
r}\nonumber\\
&+&\frac{1}{\pi r}\frac{\partial}{\partial
r}\left[(r^{1/2}-r_{out}^{1/2})\frac{\partial\dot{M}}{\partial
r}\right]\,\,\,,
\end{eqnarray}
where $r_{out}$ is the outer radius of the disk and $\nu$ is the
kinematic viscosity coefficient. The local accretion rate is given
by
\begin{eqnarray}
\dot{M}=6\pi r^{1/2}\frac{\partial}{\partial
r}(r^{1/2}\nu\Sigma)\,\,\,.
\end{eqnarray}
Equation (1) describes a disk which is continuously supplied not
only with the matter but also with angular momentum. If the mass
inflow is variable with time, the thermally stable upper branch of
the S-shaped instability ("active" phase) is characterized by an
accretion rate higher than the stationary value, while the stable
lower branch ("quiescence" phase) has a lower accretion rate than
the stationary accretion rate, and so gas is accumulated in the
outer region of the disk \citep{Sie96}. In this model, the disk is
assumed to be in the configuration of advection-dominated
inflow-outflow solutions (ADIOS)\citep{Bla99} in the lower branch,
in which most of the material transferred from the quasar host
galaxy to the accretion disk is ejected in the form of outflowing
wind. This is based on the observations that the quasar is not
active on the lower branch, which hints that the efficiency in
converting the gravitational energy to radiation is lower, thus
the material transferred from the host galaxy can be accumulated
and clumped into the outer region of the disk initially. When the
clumped mass in the outer region is transferred into the inner
region by viscosity, the central seed BH gains mass and begins to
grow. At the same time, when the surface density of the disk
increases to a limit critical value $\Sigma_A$ (Fig1), the
ionization instability occurs, the disk jumps from the lower
branch to the upper branch. In the upper branch, the gas in the
disk completely ionized and the disk restores to stable. Unlike
the lower branch, in the upper branch, the disk is highly active,
radiation from the disk reaches the highest possible luminosity,
restricted only by the Eddington limit. At this stage, the
observed galaxy is naturally in a highly active phase and the
quasar is like a bright quasars \citep{Hol96}. In order to obtain
a limit cycle behavior of ionization instability of the disk, some
stabilizing mechanism must be taken into account in the upper
branch. \cite{Jan02} proposed that a hot outflowing wind as a
result of the disk instability should develop to provide a
stabilizing mechanism in the upper branch. Energetically, the
request for such an outflow is consistent with observations
\citep{Fen99,Ree01}. Unlike the outflows of lower branch, the hot
outflowing wind of the upper branch have different properties,
which is probably produced by another mechanism, and the mass-loss
rate in the upper branch is negligible in comparing with the
accretion rate into the BH \citep{Jan02}. After an outburst, the
outer disk attempts to refill the central disk via steady inflow
and begin the next cycle. The whole cycle repeats roughly on the
viscous time of the outer disk. Because the recurrence time-scale
of the disk is longer, the individual quasar changes too slowly to
be observed in the Galaxy. It is reasonable to assume that every
quasar, as well as the growth of the BH, has already experienced a
sufficiently large number of cycles between the quiescent and
active phase in the cosmic time.

\subsection{The available gas mass accumulated into the the quasar disk}

In subsection 2.1, we analyzed the evolution of the quasar disk
along S-shaped instability and how the BH gains its mass. It is
interesting to calculate quantitatively the BH growth on three
phases in one cycle. We first estimate that how much mass from the
quasar host galaxy can be transferred and accumulated into the
outer region of the disk. Considering our assumption that the disk
in the lower branch is in the ADIOS configuration and the
mass-loss of the wind from the quasar disk in the upper branch can
be negligible \citep{Jan02}, then the transferred mass from the
host galaxy to the disk thus can be divided into two components
when quasar disk evolves in one cycle: one component is the
infalling material with mass $(\Delta M_{bh})_{acc}$, which is
also the available gas mass accumulated into the outer region of
the quasar disk around the BH ; another is outflowing material
with mass $\Delta M_{out}$ which is ejected into the ISM in which
quasar lies. To calculate $\Delta M_{out}$ and $(\Delta
M_{bh})_{acc}$, we use the pseudo-Newtonian RADIOS model proposed
by \cite{Bec01} to describe the relation between the energy
emitted into the wind and the amount of mass per unit radius per
unit time, which is
\begin{eqnarray}
\left(\frac{l_{bh}}{\dot{m}}\right)_{out}&=&\epsilon\int^{r_{out}}_{r_{in}}\frac{r^{1/2}(r+6)}{(r-2)^3}dr\nonumber\\
&\times&
\left[\int^{r_{out}}_{r_{in}}\frac{r^{1/2}(r-6)}{(r-2)^2}dr\right]^{-1}\,\,,
\end{eqnarray}
where $\epsilon$ must satisfy $max\left(\frac{\Omega_0^2}{4},
\frac{5\Omega^2_0-1}{4}\right)<\epsilon<\Omega^2_0$, which ranges
$0.025<\epsilon<0.1$ with $\Omega_0^2=0.1$, $\Omega_0$ is positive
constant, $r$ is the radius expressed in Schwarschild radii
($R_s=\frac{2GM_{bh}}{c^2}$), $l_{bh}=\frac{L_{bh}}{L_{Edd}}$,
$\dot{m}=\frac{\dot{M}_{out}}{\dot{M}_{Edd}}$, $L_{bh}$ is the
bolometric luminosity emitted by material accreting onto the BH,
and $\dot{M}_{out}$ is the mass-loss rate. $\dot{M}_{Edd}$ and
$L_{Edd}$ is the Eddington accretion rate and luminosity,
respectively. $r_{in}$ and $r_{out}$ are the inner and outer
radius of the disk, respectively. For this reference model, we
adopt $r_{in}=6$ and $r_{out} \sim 10^2$ \citep{Nar96,Nar97},
which gives the numerical value of integral in Eq (3) $\sim 0.03$.

In the case of ADIOS situation, the accumulated mass in the outer
region of the quasar disk and the outflowing mass on viscous
time-scale ($t_{vis}$) in one cycle can be estimated as
\begin{eqnarray}
(\Delta M_{bh})_{acc}&\sim&(\dot{M}_{bh})_{acc}t_{vis}\,\,\,\\
\Delta M_{out} &\sim& \dot{M}_{out}t_{vis}\,\,\,,
 \end{eqnarray}
where $(\dot{M}_{bh})_{acc}$ is the accumulated mass accretion
rate in the quasar disk around the BH. Eventually, the accumulated
material on the quasar disk should fall into the BH. Taking
$L_{bh}\sim 0.1(\dot{M}_{bh})_{acc}c^2$, and combining Eq.(3) and
Eq.(4), we obtain
\begin{eqnarray}
(\Delta M_{bh})_{acc}\sim 0.03\epsilon \dot{M}_{out}t_{vis}\,\,.
\end{eqnarray}
Eq.(6) gives the available maximum mass in the outer region of the
disk in one cycle. Let the values of $\epsilon$ varies as
$0.025<\epsilon<0.1$, we get
\begin{eqnarray}
7.5\times 10^{-4}\dot{M}_{out}t_{vis}\le (\Delta M_{bh})_{acc}\le
3\times 10^{-3}\dot{M}_{out}t_{vis}.\nonumber
\end{eqnarray}
We conclude that $(\Delta M_{bh})_{acc}$ is about $10^{-3}$ of the
total outflowing mass. Combining $\dot{M}_{ext}t_{vis}=\Delta
M_{out}+(\Delta M_{bh})_{acc}$ and Eq.(5), we thus have
\begin{eqnarray}
\Delta M_{out}&\sim &\dot{M}_{ext}t_{vis}\,\,,
\end{eqnarray}
and then
\begin{eqnarray}
(\dot{M}_{bh})_{acc}\sim 0.03\epsilon \dot{M}_{ext}\,\,\,.
\end{eqnarray}
We shall treat $\dot{M}_{ext}$ as a model parameter, which is
determined by the total stars formed in the host galaxy
\citep{Cat99}.

\subsection{The mass growth of the black hole along
the S-shaped instability in one cycle}

In this section, we study the growth of the central BH along the
S-shaped instability in one cycle. We argue that in order to
realize the next cycle of S-shaped instability, the accumulated
mass $(\Delta M_{bh})_{acc}$ of the quasar disk must be either
swallowed by the BH or ejected away in the upper branch when the
quasar is in highly active state. As the upper branch of the
S-shaped instability involves physics that is currently opened
\citep{Jan02,Bla02, Mir99}. We estimate the properties of the
outflowing and infalling in the upper branch according to the
observations. It is known that jets are among the most ubiquitous
of astrophysical phenomena, and are associated with objects
ranging from proto-stars and binary systems in our Galaxy to SMBHs
in the centers of AGNs. Large asymptotic Lorentz factors are
implied by observations of superluminal motion in many blazars
\citep{Der92,Ver94}, and the EGRET instrument on board the Compton
Gamma Ray Observatory (CGRO) has also detected intense
$\gamma$-ray flares from many of these sources
\citep{Weh98,Kat99}. The mechanisms responsible for producing the
observed jets are still poorly understood, but collimated outflows
in general appear to be associated with objects that derive their
luminosity from the accretion of matter onto a gravitating object
\citep{Der92}. The observations of relativistic jets shows that
asymptotic Lorentz factors is about $\gamma_{jet}\ge 10$
\citep{Mez92, Qui96}. For an accretion rate $\dot{M}_{bh}$, the
emitted luminosity can be estimated as
\begin{eqnarray}
L_{bh}=0.1\dot{M}_{bh}c^2\,\,\,.
\end{eqnarray}
If the outflow gas expands adiabatically, the total energy per
unit mass in the jet at a rate $\dot{M}_{jet}$ follows
\begin{eqnarray}
L_{jet}=\gamma_{jet}\dot{M}_{jet}c^2\,,\,\,\,\,\,\,.
\end{eqnarray}
It therefore follows immediately that
\begin{eqnarray}
\frac{\dot{M}_{jet}}{\dot{M}_{bh}}=\frac{1}{3\gamma_{jet}}\ll
1\,\,.
\end{eqnarray}
Eq.(11) thus shows that the mass flux in the jet on the upper
branch is only less than $3\%$ of the mass flux into the BH. This
simple energetic argument predicts that outflowing mass can be
ignored in comparing with inflowing mass in the upper branch. We
therefore neglect the mass loss in the upper branch. This means
that all accumulated mass onto the quasar disk will be swallowed
by the BH before a new cycle starts. Consequently, we have
\begin{eqnarray}
\Delta M_{bh}=(\Delta
M_{bh})_{acc}=0.03\epsilon\dot{M}_{ext}t_{vis}\,\,\,,
\end{eqnarray}
where $\Delta M_{bh}$ is the total mass growth of the BH gained in
each cycle of the S-shaped instability.

We are also interested in the detail, with observation
constraints, how the BH grows up along the S-shaped instability of
the quasar disk. It is known that the surface density and the
local accretion rate of the quasar disk evolves on the viscous
timescale $t_{vis}$ \citep{Lig74} $t_{vis}\sim R/v_R$ (where $v_R$
is the radial velocity, and $R$ is the accretion radius of the
disk), thus the life time of a quasar accretion disk in each cycle
is $t_{vis}$. The recurrence timescale $t_{current}$ of S-shaped
instability of the quasar disk is addressed as $t_{current}\sim
t_{vis}$, which operates over $10^4 $ - $10^6 \,{\rm{yr}}$
\citep{Sie96}. Time-scales spent by the quasar disk in lower,
transition and upper branches in one cycle are $t_{low}$,
$t_{tran}$ and $t_{up}$, respectively, which are given by
\cite{Sie96},
\begin{eqnarray}
 t_{low}&\sim& 0.75t_{vis}\,\,\,\\
 t_{up}&\sim& 0.1t_{vis}\,\,\,,\\
 t_{tran}&\sim& 0.15t_{vis}\,\,\,,
\end{eqnarray}

We assume that the total growth mass $\Delta M_{bh}$ of the BH in
one cycle is built up when the quasar disk passes three different
phases. Then the total growth mass of the BH in one cycle is
\begin{eqnarray}
\Delta
M_{bh}=\dot{M}_{bhl}t_{low}+\dot{M}_{bhu}t_{up}+\dot{M}_{bht}t_{tran}\,\,\,,
\end{eqnarray}
where $\dot{M}_{bhl}$, $\dot{M}_{bht}$ and $\dot{M}_{bhu}$ are the
accretion rate of the BH in lower, transition and upper branches,
respectively. Each accretion rate is limited by Eddington
accretion rate.

The accretion rate in the transition phase can be estimated from
the feature of S-shaped instability. It is noted that the
ionization instability of the quasar disk occurs when the
accretion rates of the disk are in the range $(\dot{M}_2,
\dot{M}_1)$ (see, Fig.1). We thus approximate
\begin{eqnarray}
\dot{M}_{bht}=\frac{1}{2}\left(\dot{M}_{bhl}+\dot{M}_{bhu}\right)\,\,\,,
\end{eqnarray}

Inserting Eq.(13) - Eq.(15) and Eq.(17) into Eq.(16), we get
\begin{eqnarray}
&& \Delta M_{bh}\approx 0.75 \dot{M}_{bhl}t_{vis}
 +0.1\dot{M}_{bhu}t_{vis}\nonumber\\
&&+0.075\left(\dot{M}_{bhl}+\dot{M}_{bhu}\right)t_{vis}\,\,.
\end{eqnarray}
Comparing Eq.(12) with Eq.(18), we get
\begin{eqnarray}
0.03\epsilon\dot{M}_{ext}=0.825 \dot{M}_{bhl}
 +0.175\dot{M}_{bhu}\,\,\,.
\end{eqnarray}
If $0.01\le\frac{\dot{M}_{bhl}}{\dot{M}_{bhu}}\le 0.1$
\citep{Mir99}, the accretion rates in three phase can be estimated
roughly
\begin{eqnarray}
1.64\times 10^{-3}\epsilon\dot{M}_{ext}&\le \dot{M}_{bhl}\le&1.16\times 10^{-2}\epsilon\dot{M}_{ext}\,\nonumber\\
1.16\times 10^{-1}\epsilon\dot{M}_{ext}&\le \dot{M}_{bhu}\le&
1.64\times
10^{-1}\epsilon\dot{M}_{ext}\,\,\,\nonumber\\
 2.28\times 10^{-2}\epsilon\dot{M}_{ext}&\le \dot{M}_{bht}\le&
6.38\times
10^{-2}\epsilon\dot{M}_{ext}\,\,\nonumber\\
\end{eqnarray}
If $\dot{M}_{bhu}=\dot{M}_{Edd}$, the limits of $\dot{M}_{ext}$
required by our model can be inferred from Eq.(20), which gives
\begin{eqnarray}
6 M_\odot\,{\rm yr^{-1}}&\le \dot{M}_{ext}\le& 86 M_\odot\,{\rm
yr^{-1}}\,\,\,
\end{eqnarray}
for $ \epsilon =0.1$\,\,,
\begin{eqnarray}
 24.4 M_\odot\,{\rm
yr^{-1}}&\le \dot{M}_{ext}\le& 345 M_\odot\,{\rm yr^{-1}}\,\,\,
\end{eqnarray}
for\,\, $\epsilon =0.025$\,\,.

The mass growth of the BH in three phases of one cycle can be
obtained
\begin{eqnarray}
1.23\times 10^{-3}\epsilon\dot{M}_{ext}t_{vis}&\le \Delta M_{bhl}\le& 8.7\times 10^{-3}\epsilon\dot{M}_{ext}t_{vis}\nonumber\\
1.16\times 10^{-2}\epsilon\dot{M}_{ext}t_{vis}&\le \Delta M_{bhu}\le& 1.64\times 10^{-2}\epsilon\dot{M}_{ext}t_{vis}\nonumber\\
3.42\times 10^{-3}\epsilon\dot{M}_{ext}t_{vis}&\le \Delta
M_{bht}\le& 9.57 \times
10^{-3}\epsilon\dot{M}_{ext}t_{vis}\nonumber\\
\end{eqnarray}
where $\Delta M_{bhl}=\dot{M}_{bhl}t_{low}$, $\Delta
M_{bhu}=\dot{M}_{bhu}t_{up}$, and $\Delta
M_{bht}=\dot{M}_{bht}t_{tran}$ are the mass growth of the BH in
the lower, upper and transition branches, respectively.

\section{The present mass of BH and the BH - bulge mass relation}
The total mass of the resulting BH at present is the sum of the
growth mass through the seed BH accreting in each cycle during the
Hubble time. Taking into account model assumption that the
accretion rate onto the quasar disk is limited by
Eddington-limits, we can rewrite Eq.(12)
\begin{eqnarray}
\frac{dM_{bh}}{dt} =\left\{ \begin{array} {r@{\quad;\quad}l}
 0.03\epsilon\dot{M}_{ext} & if\,\,\,
0.03\epsilon\dot{M}_{ext}\le \dot{M}_{Edd}\\
\dot{M}_{Edd} & if\,\,\, 0.03\epsilon\dot{M}_{ext}\ge
\dot{M}_{Edd}
\end{array}
\right.
\end{eqnarray}
The solutions of Eq.(24) are:

(i) If $0.03\epsilon\dot{M}_{ext}\le \dot{M}_{Edd}$,
\begin{eqnarray}
 M_{bh}(t)=0.03\epsilon\dot{M}_{ext}t+ M_{bh}(0),
 \end{eqnarray}
where t ranges $0\le t\le t_H$, $t_H$ is the Hubble timescale
($t_H=10\,{\rm Gyr}$). $M_{bh}(t)$ and $M_{bh}(0)$ are the present
BH mass and the seed BH mass, respectively.

For $0.025<\epsilon<0.1$, the seed BH mass $M_{bh}(0)$ are:
\begin{eqnarray}
2\times 10^6M_\odot\le M_{bh}(0)\le 3\times 10^7M_\odot
\end{eqnarray}
Then the present BH mass becomes $2\times 10^8 M_\odot$ and
$3\times 10^9M_\odot$ for the seed BH mass $2\times 10^6 M_\odot$
and $3\times 10^7M_\odot$ respectively.

(ii) If $ 0.03\epsilon\dot{M}_{ext}\ge \dot{M}_{Edd}$ \,\,\,\,,
\begin{eqnarray}
M_{bh}(t)=M_{bh}(0)e^{(t/t_{Edd})}\,\,\,,
\end{eqnarray}
where $t_{Edd}$ is Eddington accretion time. This solution is not
important because the BH will first grow exponentially, quickly $
0.03\epsilon\dot{M}_{ext}\le \dot{M}_{Edd}(M_{bh}(t))$ it switches
back to the first solution ( Eq.(25)). Since the time spends in
the exponential growth is less than the Hubble time so this stage
can be ignored unless the seed BH mass is extremely small.
Consequently, the solutions of Eq.(24) suggest that a typical
quasar BH with a final mass $2\times 10^8M_{\odot} $ can be grown
from a seed BH with mass $2\times 10^6M_{\odot} $.

As assumed that the material transferred from the host galaxy is
proportional to the mass that has formed stars, this situation is
strictly true in model (i) of \cite{Cat99}, then $\Delta
M_{out}\approx\dot{M}_{ext}t_{vis}{}\propto M_{bulge}$, where $
M_{bulge}$ is the mass of host galactic bulge. Taking into account
of Eq.(7) and Eq.(12), and let $0.025\le\epsilon \le 0.1 $, we
thus obtain the mass ratio of the BH to the galactic bulge
\begin{eqnarray}
7.5\times 10^{-4} \le \frac{M_{bh}}{M_{bulge}}\le 3\times 10^{-3}
\,\,\,\,.
\end{eqnarray}

\section{Discussion and conclusion}
A quasar disk model associated with the ionization instability is
proposed to account for the build up of SMBHs in quasar host
galaxies. The growth of the BH is studied when the quasar disk
evolves along S-shaped instability. We shows that the main growth
of the BH occurs in the upper branch when quasar disk is in a
highly active state. Two model parameters $\epsilon$ and
$\dot{M}_{ext}$ are involved. Though in quasar accretion disks the
gas supply mechanism is unknown \citep{Sie96}, the limits for an
available gas supply with $ 6 M_\odot\,{\rm yr^{-1}}\le
\dot{M}_{ext}\le 86 M_\odot\,{\rm yr^{-1}}$ from the quasar host
galaxy is deduced based on observations and the limits of
Eddington-accretion rate. Our conclusion can be summarized as: (i)
We first demonstrate that the BHs of the quasar host galaxies can
grow up through accretion with two model parameters. With
$0.025<\epsilon<0.1$, a BH with mass $2\times 10^8 M_\odot$ can
grow up from a seed BH with mass $2\times 10^6 M_\odot$. (ii)
Secondly, the observed BH - bulge mass relation \citep{Mag98} can
be inferred, which is $7.5\times 10^{-4} \le
\frac{M_{bh}}{M_{bulge}}\le 3 \times 10^{-3}$ in our model
parameters. (iii) Furthermore, in present scheme, the massive seed
black hole is required in order to build up super-massive BH
through accretion if the gas is provided from the quasar host
galaxy. This hints that a low-mass stellar BH may not grow to
super-massive black hole unless without Eddington-limit. The
details for the formation of massive seed BH ($\sim 10^6 M_\odot$)
and if there is a sufficient cold gas supplied to the BH are
beyond the scope of this paper.

\acknowledgments

We are grateful to Dr. Y. F. Huang and Dr. X. Z. Zheng for helpful
discussions. This work was supported by a RGC grant of the Hong
Kong SAR government, the National Natural Science Foundation of
China (NSFC 10273011), the National 973 Project (NKBRSF
G19990754), and the Special Funds for Major State Basic Research
Projects.

\clearpage



\clearpage

\end{document}